\documentclass[letter]{aa}  
\nolinenumbers

\usepackage{titlesec}
\usepackage{dblfloatfix}
\usepackage{graphicx}
\usepackage{txfonts}
\usepackage{CJKutf8}
\usepackage{lipsum}
\usepackage{subcaption}         % necessary for continued figures, example in section 3
\usepackage{lscape}             % to rotate a single page table, example in appendix.
\usepackage{placeins}           % useful with \FloatBarrier, to keep 
\newcommand{\CNnames}[1]{{\begin{CJK}{UTF8}{gbsn}~(#1)~\end{CJK}}}
\newcommand{\megamite}{KIC\,2309579}
\newcommand{\kepler}{\textit{Kepler}}
\newcommand{\dKa}{\ensuremath{\delta K_{\rm a}}}
\newcommand{\muHz}{\ensuremath{\mathrm{\mu Hz}}}
\newcommand{\dKth}{\ensuremath{\delta K_{\rm Th}}}
\newcommand{\nurot}{\ensuremath{\nu_{\rm rot}}}
\newcommand{\jb}[1]{#1}

\newcommand{\rev}[1]{#1}

\begin{document}
\nolinenumbers

   \title{Seismic signature of a magnetic field in the $\gamma$ Doradus star \megamite}

   %\subtitle{Subtitle}

%%%%%%%%%%%%%%%%%%%%%%%%%%%%%%%%%%%%%%%%
% Please do not include ORCIDs next to author names.
% Only ORCIDs authenticated by individual authors in EDP Sciences editorial system will be taken into account.
% ORCIDs included here will be removed.
%%%%%%%%%%%%%%%%%%%%%%%%%%%%%%%%%%%%%%%%

   \author{S. Ihallaine\inst{1}
        \and J. Ballot\inst{1} %\fnmsep\thanks{Shows the usage of elements in the author field}
        \and F. Lignières\inst{1}
        \and L. Ferrié\inst{1}
        \and S. Charpinet\inst{1}
        \and M. Galoy\inst{2}
        \and G. Li\CNnames{李刚}\inst{3}
        }

   \institute{IRAP, Université de Toulouse, CNRS, CNES, 14 avenue Edouard Belin 31400 Toulouse, France\\
             \email{selyan.ihallaine@utoulouse.fr}
            \and 
            Max-Planck-Institut für Sonnensystemforschung, 37077 Göttingen, Germany
            \and
            Centre for Astrophysics, University of Southern Queensland, Toowoomba, QLD 4350, Australia}

   \date{Received 16 April 2026 / Accepted 21 May 2026}

% \abstract{}{}{}{}{}
% 5 {} token are mandatory
 
  \abstract
  % context heading (optional) leave it empty if necessary  
   {Internal magnetic fields have recently been detected and measured in the radiative core of red giant stars using asteroseismology. Being one of red giant stars progenitors and exhibiting high radial order gravity modes, $\gamma$ Doradus stars are also good candidates to hold detectable magnetic fields in their radiative envelope.}
   {We aim to detect internal magnetic field in a rapidly rotating $\gamma$ Doradus star for the first time, through its influence on the propagation of Kelvin gravito-inertial modes. } 
   {We used the seismic variable $\dKa$, defined as a combination of Kelvin modes frequencies, which is sensible to the presence of a magnetic field. Following the detection, we carried out a modelling of the star oscillation spectrum considering a magnetic component following a Bayesian approach.}
   {We found a magnetic signature into the radiative envelope of \megamite. If located just above the core, in the layers that were previously convective, the magnetic field would reach $\sim$ 4 kG.}
   {}


   \maketitle
\nolinenumbers

%%%%%%%%%%%%%%%%%%%%%%%%%%%%%%%%%%%%%%%%%%%%%%%%%%%%%%%%%%%%%%
\section{Introduction}
Over the past few decades, asteroseismology has emerged as a powerful tool for probing the interior of stars and has revealed crucial information like near-core rotations, ages and masses. These measurements have shown that stellar models over-estimate the core rotation \citep{Mosser2012, Ouazzani2019}, which leads to the point that we miss out some physical phenomenon occurring in stars. Magnetic fields
may be a major actor of the angular momentum transport within stars. Nevertheless, for a long time, only surface magnetic fields could be detected and studied using spectropolarimetry, leaving the study of the inner part inaccessible. 
\citet{Fuller2015} have shown that a strong enough inner magnetic field hinders the propagation of gravity (g) modes and could explain the suppression of mixed dipolar modes observed in some red giants stars. 
More recently, magnetic fields have been detected and measured in the radiative core of several red giants stars \citep{Li2022, Li2023, Deheuvels2023, Hatt2024, Villate2026} and in the radiative envelope of a slowly rotating $\delta$ Scuti-$\gamma$ Doradus star \citep{Takata2026} using a seismic approach. These measurements provide key constraints for answering the question of angular momentum transport.
The $\gamma$ Doradus ($\gamma$ Dor) stars are pulsating main-sequence (MS) late A- to early F-type stars with masses from 1.3 to 2\:M$_\sun$ 
\citep{Kaye1999}. Their internal structure is composed of a convective core, a radiative envelope and a shallow convective envelope. Although they exhibit high radial order gravito-inertial (g-i) modes, which are most affected by magnetic fields, 
there is currently no seismic evidence of inner magnetic field in the typical rapidly rotating $\gamma$ Dor stars. 
Their rotation periods, of the order of a day \citep{Li2020}, makes it impossible to consider rotation as a perturbation as it is done for red giants \citep{Li2022}. The traditional approximation of rotation \citep{Lee1997}, hereafter named TAR, has been used instead to study g-i modes of rapidly rotating stars, as its validity has been tested for the modes observed in $\gamma$ Dor stars \citep{Ballot2012,Ouazzani2017,Ouazzani2020}. \citet{Prat2019, Prat2020} studied the effect of dipolar magnetic fields on TAR g-i modes considering the Lorentz force as a perturbation while \citet{Dhouib2022} treated the Lorentz force in a non-perturbative way for an axisymetric toroidal field. Recently, \citet{Lignieres2024} proposed a generalization to an arbitrary magnetic field of the perturbative approach and derived analytical forms of frequency shifts for g-i and Rossby modes. They also introduced a new seismic variable, named $\dKa$ for difference of Kelvin mode frequencies, which is specifically sensitive to magnetic fields.

In the following, we first present the method for identifying magnetic star candidates thanks to the seismic variable \dKa\ and carry out the spectral analysis of such a star, \megamite, in Sect. \ref{sec:Candidates}. We then present the modelling of the star oscillation spectrum considering a magnetic component and the corresponding results in Sect. \ref{ModelMag}. In Sect. \ref{MagFieldProp}, we investigate the properties of the inner magnetic field of \megamite. We finally propose a discussion and a conclusion in Sect. \ref{Conclu}.
%%%%%%%%%%%%%%%%%%%%%%%%%%%%%%%%%%%%%%%%%%%%%%%%%%%%%%%%%%%%%%
\section{Identifying magnetic candidates}\label{sec:Candidates}
Kelvin modes are g-i equatorial prograde modes present in rotating stars. They are usually labelled by their degree $\ell$ and their azimuthal order $m=-\ell$ as they become prograde sectoral modes at zero rotation.
\citet{Lignieres2024} have introduced a seismic variable, \dKa, which is defined as a combination between the frequencies of $\ell=1$ and $\ell=2$  Kelvin modes with the same radial order $n$,  denoted $\nu_{n,\ell}$, normalised by twice the star rotation frequency \nurot:
\begin{equation}
    \dKa = \dfrac{1}{2 \nurot}  \left( \nu_{n,1} - \dfrac{1}{2} \ \nu_{n, 2} \right).
\end{equation}
In the absence of magnetic field, asymptotic theory predicts that \dKa\ is the same function of the spin parameter $s_{n,1}=2 \nurot/\nu^{\rm co}_{n,1}$, where $\nu^{\rm co}_{n,1}$ refers to the $\nu_{n,\ell}$ frequency in the co-rotation frame, for all $\gamma$~Dor stars and that this function vanishes at high $s_{n,1}$. In contrast, the radial component of a magnetic field produces an increase of \dKa\ proportional to $s_{n,1}^3$. This property provides a simple method for detecting magnetic fields in $\gamma$ Dor stars holding both $\ell=1$ and $\ell=2$ Kelvin modes.

\subsection{Description of the analysed $\gamma$ Dor star sample}\label{ssec:sampledescription}
\begin{figure}[!htbp]
    \centering
    \includegraphics[width=0.85\linewidth]{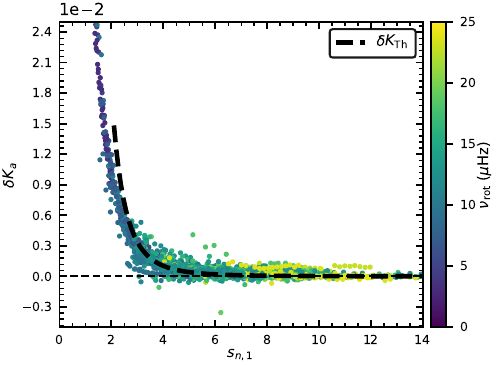}
    \caption{Dimensionless seismic variable \dKa\ as a function of the spin parameter $s_{n,1}$ for the 155 $\gamma$ Dor stars from the \citet{Li2020} catalogue. A black dashed curve shows the asymptotic approximation \dKth\ \citep[see][]{Lignieres2024}. Colors indicate the stellar rotation rate.}
    \label{fig:dKa_allstars}
\end{figure}

\citet{Li2020} analysed 611 $\gamma$ Dor stars observed by the \textit{Kepler} space mission \citep{Borucki2010}. They found both $\ell=1$ and $\ell=2$ Kelvin modes among 155 of these stars. From a $\Delta P - P$ diagram fitting method, they determined the rotation frequencies and the buoyancy radii $\Pi_0$ of these stars. We identified the radial order of modes detected in these stars with the TAR and computed \dKa\ for this sample. Results are shown in Fig.~\ref{fig:dKa_allstars}. The data collapse around a single curve, independent of the rotation rate of the stars, that converges at high spin parameters towards the asymptotic relation calculated within the TAR and the Wentzel-Kramers-Brillouin (WKB) approximation, denoted \dKth.
We thus searched for stars showing deviations from this general trend.
Some points exhibit strong departures, but we have shown it is  generally due to incorrect radial order identifications. We identified a few stars with a behaviour compatible with what we expect from a magnetic field.
In the rest of the paper, we present the detailed analysis of the best candidate, \megamite.

\subsection{Spectral analysis of \megamite}
In order to fully control all the steps of the analysis, we first reprocessed the \kepler\ light curve of \megamite\ downloaded from the MAST archive\footnote{\url{https://archive.stsci.edu/missions-and-data/kepler}} to extract the oscillation frequencies using the \texttt{FELIX} code \citep{Charpinet2010, Zong2016}. 
We extract the frequencies with a signal-to-noise ratio (SNR) greater than 4.7, that is a 4-$\sigma$ detection level.
The detailed result of the extraction is reported in Appendix~\ref{app:felix}. As in \citet{Li2020}, we identified $\ell=1$ and $\ell=2$ Kelvin modes in the ranges 13--17\:\muHz\ and 28--32\:\muHz. The mode identification is performed with the \texttt{GMorse} code \citep{Christophe2018}.

\begin{figure}[!htbp]
    \centering
    \includegraphics[width=0.85\linewidth]{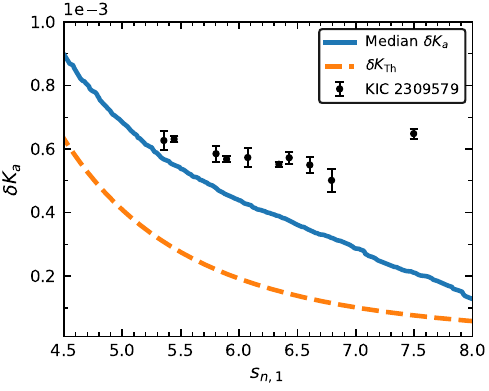}
    \caption{Dimensionless seismic variable \dKa\ as a function of the spin parameter $s_{n,1}$ for modes identified in \megamite. For comparison, we plot the median \dKa\ of the 155 stars (blue line) and the asymptotic approximation \dKth\ (orange dashes).}
    \label{dKa_2309579}
\end{figure}

We excluded from the following analysis the frequencies that have high probabilities ($\leq 2 \sigma$) to be linear combinations \rev{of two higher amplitude frequencies} or harmonics of higher peaks. Figure~\ref{dKa_2309579} shows \dKa\ of \megamite\ computed from this frequency list. For comparison, are also plotted the median \dKa\ of the 155 stars and \dKth.
The values measured for \megamite\ exhibit a clear departure from the median of the whole sample and from \dKth. Furthermore, we found that this departure is compatible with a profile proportional to $s_{n,1}^3$ as expected for magnetic stars. \rev{The difference between \dKth \ and the median \dKa \ can be partly explained by the limitations of the asymptotic theory \citep{Lignieres2024} but it could also result from a general magnetic behaviour of the stars sample. The small oscillatory behaviour of the data is likely due to a glitch occurring near the convective core of the star.}
The variable \dKa\ is a good indicator for identifying candidates, but shows limitations: (i) results rely on an \emph{a priori} determination of $n$, which can be biased by the presence of a magnetic field \citep{Lignieres2024} (ii) we only used 20 out of the 34 extracted frequencies (those that sharing a common $n$ and different $\ell$).
In the following, we then propose to use a magnetic model to reproduce \megamite\ oscillation spectrum.

\section{Spectrum model including magnetic field}\label{ModelMag}
Following an approach similar to the one used to analyse red giant stars by \citet{Villate2026}, we fit all the observed mode frequencies with an asymptotic model including the effects of a magnetic field.
We present the model in Sect.~\ref{Description} and  the results of the fit to the observations in Sect. \ref{results}.

\subsection{Asymptotic model}\label{Description}
In the absence of a magnetic field, we compute the frequencies of Kelvin modes within the TAR and using the WKB approximation in the radial direction. 
In this case, the non-perturbed frequencies $\nu_{n, \ell}^0$ of modes are expressed through the equation:
\begin{equation}
    \frac{1}{\nu_{n, \ell}^0 - m \nu_{\text{rot}}} = \dfrac{\Pi_0}{\sqrt{\Lambda_{\ell} \left(s_{n,\ell}\right)}} \left(n + \epsilon_{g_\ell} \right),
\end{equation}
where $\Lambda_{\ell}\left(s\right)$ are the Laplace tidal equation eigenvalues, $\Pi_0$ is the buoyancy radius of the star, and $\epsilon_{g_\ell}$ is a phase offset for modes of degree $\ell$ \jb{\citep[e.g.][]{Berthomieu1978}}. \citet{Lignieres2024} derived the perturbation of frequencies induced by a magnetic field non dominated by its azimuthal component for $\ell = 1$ and $\ell = 2$ Kelvin modes:

\begin{equation} \label{perturb_l1}
    \nu_{n, 1} = \nu_{n, 1}^0 + \nu_B \cdot s_{n, 1}^3, \quad \nu_{n, 2} = \nu_{n, 2}^0 + 4 \nu_B \cdot s_{n, 2}^3,
\end{equation}
where $\nu_B = \dfrac{\mathcal{I} B_{\text{eq}}^2}{(4\pi)^5 \nurot^3}$ in cgs units, with
\begin{equation}
  \mathcal{I} = \left. \int_{r_i}^{r_o} \frac{1}{\rho} \ \left( \frac{N}{r} \right)^3 \,\text{d}r \, \middle/ \int_{r_i}^{r_o} \frac{N}{r} \,\text{d}r\right.\quad\mbox{and}
\end{equation}
\begin{equation}\label{B_eq}
    B_{\text{eq}}^2 = \int_{r_i}^{r_o} K_r(r) \ \langle B_r^2 \rangle_{\phi}\left(\theta = \pi/2\right) \ \text{d}r,
\end{equation}
where $\rho$ is the density, $N$ the Brunt-Väisälä frequency, and $r_i$ and $r_o$ are respectively the inner and outer boundaries of the oscillation cavity. The factor $\mathcal{I}$ depends on the stellar structure along the oscillation cavity and $B_{\text{eq}}^2$ is the radial and azimuthal average of the radial component of the magnetic field evaluated at the equator\rev{, where Kelvin modes are mainly concentrated for large spin parameters}. The radial weight function $K_r(r)$ is defined as
\begin{equation}
    K_r(r) = \left.\dfrac{1}{\rho} \ \left( \dfrac{N}{r} \right)^3 \ \middle/ \displaystyle \int_{r_i}^{r_o} \dfrac{1}{\rho} \ \left( \dfrac{N}{r} \right)^3 \ \text{d}r\right..
\end{equation}
Compared to Eqs. 30 and 31 in \citet{Lignieres2024}, we here leave out a $\propto 1/s$ term that appears to be negligible in front of the first order terms. The model relies then on five free parameters: $\Pi_0, \ \nu_{\text{rot}}, \ \epsilon_{g_1}, \ \epsilon_{g_2}$ and $\nu_B$.

\begin{figure}[!htbp]
    \centering
    \includegraphics[width=0.93\linewidth]{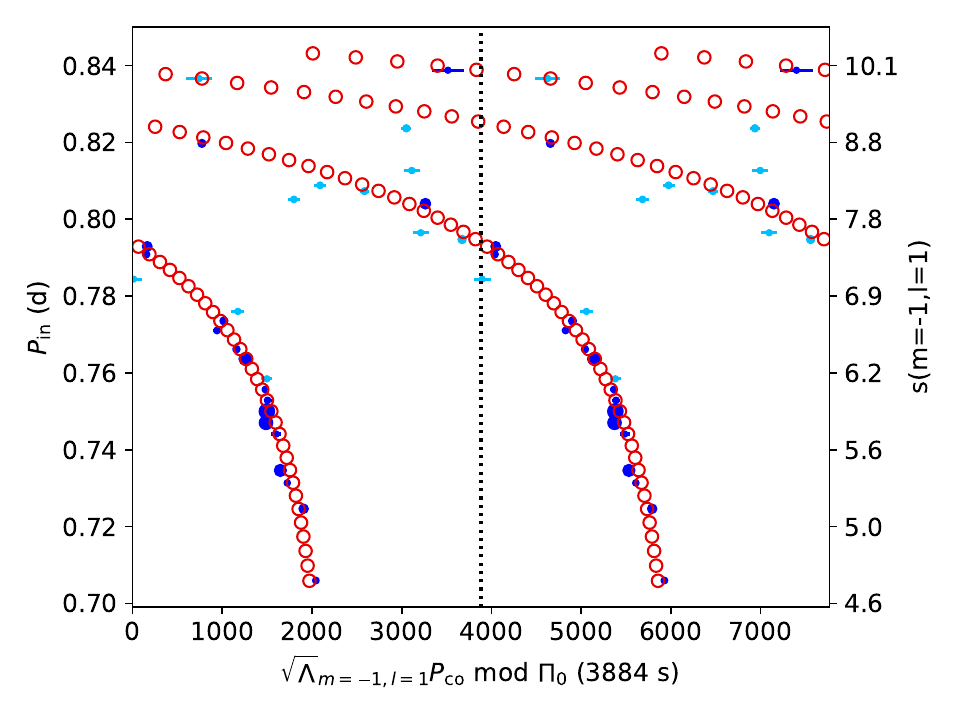}
    \includegraphics[width=0.93\linewidth]{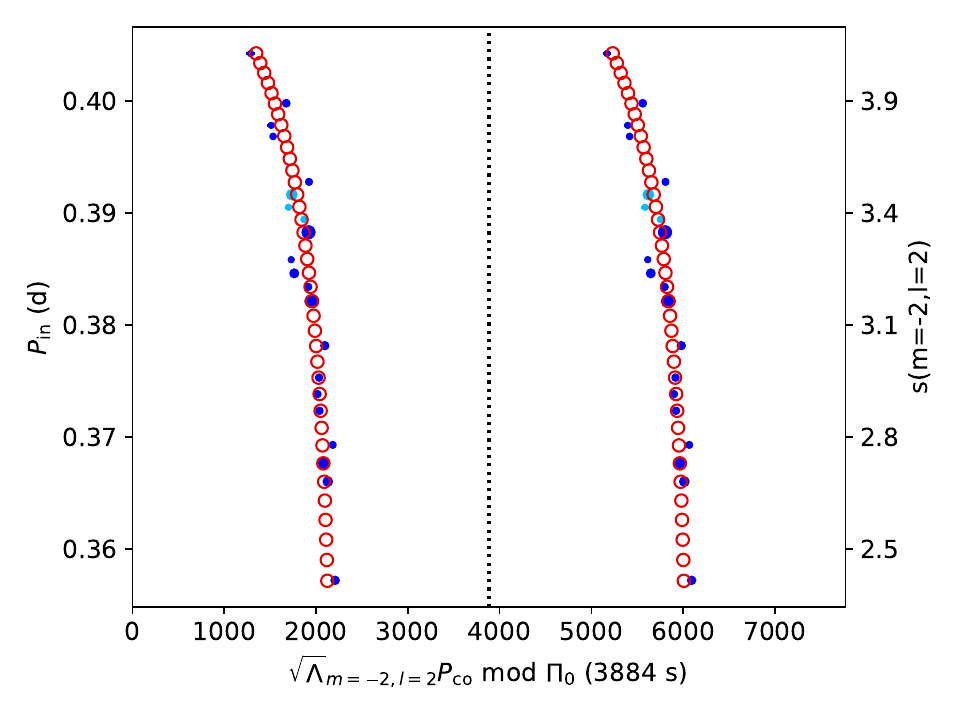}
    \caption{\jb{Stretched period} \'echelle diagram of $\ell=1$ (top) and $\ell=2$ (bottom) Kelvin mode frequencies. Navy blue dots: observed frequencies used for the analysis. Light blue dots: observed frequencies that are possible frequency combinations and discarded for the analysis. Red circles: frequencies of the best model including magnetic field. The size of the dots is proportional to the mode amplitude.}
    \label{fig:echelle_l12_mag}
\end{figure}

\subsection{Fitting the model to observations}\label{results}

We fit our model to the observed frequencies by following a Bayesian approach using the \texttt{ABIM} code (see Appendix \ref{ABIM} for details).
We obtained unimodal (almost normal) posterior distributions for all parameters. We compared the best-model frequencies, obtained from the median values of these distributions, to the observed ones. We plot the comparison in a so-called stretched period \'echelle diagram \citep{Christophe2018} in Fig.~\ref{fig:echelle_l12_mag}. The best model is consistent with the observations.
In a non-magnetic TAR framework, the échelle diagrams represent mode periods sharing a same $\ell$ and $m$ on a vertical line. 
In our case, the échelle diagrams, especially that of the $\ell=1$ modes, show distinct curvatures due to the influence of the magnetic field on the higher periods. We also plotted in light blue the frequencies we excluded from the analysis because they could be interpreted as frequency combinations. Some of them could nevertheless be modes since they are very \rev{close} to modeled frequencies.

We then derived the posterior distribution of $\mathcal{I}B_{\text{eq}}^2$ from those of $\nu_B$ and $\nurot$. We thus measured that $\mathcal{I}B_{\text{eq}}^2 = 2.23\pm 0.16 \times 10^{-20}\:\text{cm}\:\text{s}^{-2} \text{g}^{-1} \text{G}^2$. In order to infer a magnetic field strength, we need a stellar structure model of \megamite.

\section{Magnetic field strength}\label{MagFieldProp}

We computed $\gamma$ Dor star models with \texttt{CESAM2k20} \citep{Morel1997, Manchon2025} calibrated on the buoyancy radius of \megamite\ that we have measured, its luminosity, effective temperature and metallicity observed by \textit{Gaia} \citep{Gaia2023}. We consider different mixing lengths in the convective envelope and different overshoot and mixing prescriptions above the core (see Appendix~\ref{app:cesam}). The weight function $K_r$ peaks in two regions in our models: above the convective core, where composition gradients are strong, and below the convective envelope (see Fig.~\ref{K_r}). From one model to another, the relative contribution of these two regions may change significantly. Indeed, when the convective envelope is shallower, the weight of the upper layers increases. As a consequence $\cal I$ varies from one model to another, and the measured magnetic field averaged over the radiative region ranges from 140 to 590\:G. To go further, since $K_r(r)$ is bimodal, we need to consider two possibilities: either (i) the field is located below the convective envelope, or (ii) it is located above the core. 

In the first case, an average magnetic field $B_r \approx 140-700$\:G below the envelope reproduces the observed signature. However, such a field appears to be larger by an order of magnitude than the critical field in this region (see an example in Fig.~\ref{Bcrit}). Such a magnetic field would be strong enough to suppress g modes \citep{Fuller2015}. We thus must rule out this possibility.

For the second scenario, we assume that a magnetic field is present in the layers above the core that were previously convective. This field is thought to be a remnant of the convective core dynamo-generated field, left behind as the convective core retreats. In this case, a field \rev{$B_r = 4 \pm 0.3$\:kG} 
reproduces the observed signature. \rev{The error bar reflects the dependence on} the star models of \megamite. Moreover, we verified that it is below the critical field for all the star models. It thus clearly favours this second scenario.

\section{Discussion and conclusion}\label{Conclu}

If the presence of a magnetic field perfectly explains the spectrum of \megamite, we must verify that other phenomena cannot mimic the signature of a magnetic field. We especially considered the presence of glitches, differential rotation, or dips induced by pure inertial modes in the core.

First, glitches are expected to occur in $\gamma$ Dor stars due to the sharp increase of the Brunt-Väisälä frequency above the convective core. However, following \citet{Miglio2008}, the frequency shift induced by a glitch for a $\ell=2$ Kelvin mode is twice the shift for the $\ell=1$ Kelvin mode with the same radial order. As a consequence, glitches \rev{tend} to vanish in \dKa\ (see Fig. \ref{dKa_2309579}) and cannot explain the observed signature.

Second, the effects of radial differential rotation have been studied by \citet{Takata2020}. They showed it can be treated within the TAR considering a rotation that is an average over the mode cavity, which is the same for $\ell=1$ and 2 Kelvin modes. Moreover, since the latitudinal variation of $\ell=1$ and $\ell=2$ Kelvin modes are very similar \citep[see, e.g.,][]{Lignieres2024}, we expect that latitudinal differential rotation affects similarly $\ell=1$ and $\ell=2$ Kelvin modes and cancel out in \dKa.

Finally, in the range of spin parameters reached in this study, $\ell=1$ Kelvin modes could couple with a purely inertial mode propagating in the convective core, creating a so-called dip \citep{Ouazzani2020}. This induces frequency shifts of modes, creating a feature in $\dKa$ that could mimic the one produced by a magnetic field. We thus decided to model the spectrum of \megamite\ by considering the presence of a dip, following \citet{Tokuno2022} and \citet{Galoy2024}. Results are presented in Appendix \ref{dip_description}. This model struggles to reproduce both $\ell=1$ and $\ell=2$ long-period modes whereas the magnetic model well explain the whole spectra (see Fig.~\ref{echelle_l12_dip}). Moreover, the coupling factor $q$ between the pure inertial mode and g-i modes is large, reaching values that are expected for early-MS stars \citep{Galoy2024} whereas \megamite\ is a mid-MS star according to our models and the models of \citet{Fritzewski2024}. In addition, the spin parameter of the inertial mode $s_\star$ has been shifted to low values compared to the expected ones \citep{Galoy2024}. To explain this, we could invoke a fast rotating core or the presence of a strong magnetic field inside the core \citep{Barrault2025A,Barrault2025B}. However, in such a case, a decrease of $s_\star$ must go with a strong decrease of $q$, in contradiction with the observation. We conclude that a dip hardly reproduces the spectrum of \megamite.

We thus detected a near-core magnetic field of $B_r \approx 4$\:kG in \megamite. It is the first seismic evidence of a magnetic field in a typical $\gamma$ Dor star as \citet{Takata2026} detection concerns a star that rotates abnormally slowly for a $\gamma$ Dor star. \citet{Li2022} proposed that magnetic fields observed in the core of red giants were remnants of core magnetic fields of the MS. 
%The field strength measured in \megamite\ is in good agreement with this scenario.
\rev{The field strength extrapolated by \citet{Li2022} for MS stars is similar to the one found in this paper and in \citet{Takata2026}, which is compatible with a convective core dynamo field.} The technique we used to analyse this star will be extended to other $\gamma$ Dor stars, observed by spaced-based missions \kepler, TESS and, in a very near future, PLATO.

%%%%%%%%%%%%%%%%%%%%%%%%%%%%%%%%%%%%%%%%%%%%%%%%%%%%%%%%%%%%%%
\begin{acknowledgements}
\rev{We thank the anonymous referee for providing helpful and constructive comments}. This work has been supported by CNES, focused on the preparation of the PLATO mission. We thanks M. Deal for his useful advice on stellar modeling, O. Creevey for her precious insight on Gaia measurements, and V. Antoci for useful discussion on spectral analysis.
This paper includes data collected by the \kepler\ mission and obtained from the MAST data archive at the Space Telescope Science Institute (STScI). Funding
for the  \kepler\ mission is provided by the NASA Science Mission Directorate.
STScI is operated by the Association of Universities for Research in Astronomy,
Inc., under NASA contract NAS 5–26555.
\end{acknowledgements}

\bibliographystyle{aa} 
\bibliography{biblio}

%%%%%%%%%%%%%%%%%%%%%%%%%%%%%%%%%%%%%%%%%%%%%%%%%%%%%%%%%%%%%%%
% Appendices must be placed after   \end{thebibliography}
% They will be placed automatically on a new page.
%%%%%%%%%%%%%%%%%%%%%%%%%%%%%%%%%%%%%%%%%%%%%%%%%%%%%%%%%%%%%%%
\begin{appendix}
%%%%%%%%%%%%%%%%%%%%%%%%%%%%%%%%%%%%%%%%%%%%%%%%%%%%%%%%%%%%%%%
% In the PDF output, floats should be placed
% under their own appendix, not before the title, nor after the
% title of the next appendix.

% In short appendices, onecolumn floats (\figure*
% or \table*) will generate a blank page.
% To prevent this behaviour, a few examples are provided here. 

% In case you have a lot of floating objects for little text and the 
% LaTeX engine moves the floats away from their context, the command
% \FloatBarrier of the “placeins” package will empty the
% float buffer and place all stored floats in the continuity.

% If you still encounter problems with wide floats placement,
% just use the onecolumn environment throughout the appendices.
%%%%%%%%%%%%%%%%%%%%%%%%%%%%%%%%%%%%%%%%%%%%%%%%%%%%%%%%%%%%%%%

%____________________________________________________________
%       Wide floats at the start of an appendix: first method
%-------------------------------------------------------------
% To prevent a blank page after the start of an appendix:
% - Switch to one \onecolumn first
% - Declare the section title
% - Declare the onecolumn float with the parameter [h!]
% - Revert to \twocolumn at the end of the section

\section{Magnetic model spectrum}

\subsection{Asymptotic magnetic model}

As described in Sect.~\ref{Description}, we modeled the spectrum of \megamite\ within an asymptotic approximation of the TAR.
%\citet{Ballot2012} and 
\citet{Christophe2018} have shown that the asymptotic TAR is well suited to model g-i mode spectra of rotating stars. Of course, there are some deviations due to non-asymptotic effects and we must ensure that these deviations do not introduce biases that could lead to false magnetic detections.
We then used spectra computed with the 2D oscillation code \texttt{TOP} \citep{Reese2006} that was used in \citet{Lignieres2024}, as well as spectra computed with the non-asymptotic TAR oscillation code \texttt{GYRE} \citep{Townsend2013}. We computed spectra of a $\gamma$ Dor star model with a mass of 1.5 M$_{\odot}$ and a radius 2.2 R$_{\odot}$ with a rotation of 11.5\:$\mu$Hz. We then fitted our asymptotic model of $\ell=1$ and $\ell=2$ Kelvin modes to these synthetic spectra. We found good agreement. We are able to fit simultaneously both series of modes with a common $\Pi_0$ and $\nurot$ by allowing two different values for the phase offset $\epsilon_g$. Values of $\epsilon_{g,1}$ and $\epsilon_{g,2}$ are very close but need to be slightly different to correctly reproduce the spectra. We then added $\nu_B$ as a free parameter in the asymptotic model. We fitted our synthetic spectra and verified we recover a value of $\nu_B$ compatible with zero.

Similarly to \citet{Villate2026},
we decided to introduce a model error, which is quadratically added to the observation errors, to take into account the dispersion of frequencies around the asymptotic model due to non-asymptotic effects.
Based on a comparison of asymptotic models to complete computations, we have considered that, for each series of modes, the error of  periods is constant in the co-rotation frame. We thus added two additional free parameters, $\sigma_{\ell_1}$ and  $\sigma_{\ell_2}$, which are the model error of $\ell=1$ and $\ell=2$ Kelvin modes.

\begin{table}[!htbp]
\caption{Prior distributions for the magnetic model}
\label{mag_seismic_prior}
\centering
\begin{tabular}{c c c}
\hline  \hline
    Parameter & Distribution & Interval \\
\hline
    $\Pi_0$ (s) & Uniform & $\left[3000, \ 5000\right]$\\
    $\nu_R \ \left( \mu \text{Hz} \right)$ & Uniform & $\left[11, \ 12 \right]$\\
    $\epsilon_{g_\ell}$ & Uniform periodic & $\left[0, \ 1\right]$ \\
    $\nu_B \ \left( \mu \text{Hz} \right)$ & Modified Jeffrey & \jb{$\left[0, 10^{-4}\right]$} \\
    $\sigma_{\ell_1}$ (s) & Uniform & $\left[0, \ 2000\right]$\\
    $\sigma_{\ell_2}$ (s) & Uniform & $\left[0, \ 1000\right]$ \\
\hline
\end{tabular}
\end{table}

\begin{table*}[!htbp]
\caption{Estimated seismic parameters for the magnetic model}
\label{mag_seismic_param}
\centering
\begin{tabular}{c c c c c c c c}
\hline  \hline
    $\Pi_0$ (s) & $\nu_R \ \left( \mu \text{Hz} \right)$ & $\epsilon_{g_1}$ & $\epsilon_{g_2}$ & $\nu_B$ (pHz) & $\sigma_{\ell_1}$ (s) & $\sigma_{\ell_2}$ (s)\\ [0.5ex]
\hline
    $3884 \pm 31$ & $11.501 \pm 0.015$ & $0.562 \pm 0.282$ & $0.563 \pm 0.275$ & $46.8 \pm 3.4$ & $76^{+28}_{-19}$ & $46^{+13}_{-9}$ \\
\hline
\end{tabular}
\end{table*}

\subsection{Fitting the observations}\label{ABIM}
We fit our model to the observed frequencies by following a Bayesian approach with a Markov Chain Monte Carlo (MCMC) method. We defined priors for all parameters of the model. The prior for buoyancy radius $\Pi_0$ follows a uniform distribution over $\left[3000, \ 5000\right]$ seconds, which covers the typical $\Pi_0$ values for $\gamma$ Dor stars \citep{Li2020}. \citet{Li2020} also showed that the seismic rotation period remains quite close to the surface one for $\gamma$ Dor stars. As \megamite\ exhibits a rotational modulation at 11.50\:$\mu$Hz (see Appendix~\ref{app:felix}), we set a uniform prior distribution over $\left[11, \ 12 \right] \ \mu$Hz for $\nu_{\text{rot}}$. We set a uniform-periodic prior distribution over $\left[0, \ 1\right]$ for the $\epsilon_{g_1}$ and $\epsilon_{g_2}$ parameters. Here, `uniform-periodic' means a uniform prior for which the posteriors are the results of steps by the walkers modulo the interval range. As we do not have prior information about the magnetic component $\nu_B$, we used a non-informative distribution, the modified Jeffrey distribution.  This prior is a scale-invariant prior over $]10^{-5}, 10^{-4}]\:\mu$Hz, and smoothly transitions to a quasi-uniform distribution over $[0,10^{-5}]\:\mu$Hz. Finally, the prior distributions for the model errors $\sigma_{\ell_1}$ and $\sigma_{\ell_2}$ are uniform over respectively $\left[0, \ 2000\right]$ and $\left[0, \ 1000\right]\:\mu$Hz for $\ell=1$ and $\ell=2$ modes. The chosen prior distributions of the model parameters are summarised in Table \ref{mag_seismic_prior}.

We used the Asteroseismic Bayesian Inference by MCMC (\texttt{ABIM}) code to sample the posterior probabilities. \texttt{ABIM} is a Fortran code parallelised with OpenMP directives, which we have developed. It has been used for example to model mixed mode spectra in red giants in \citet{Villate2026}. For the present work, sampling has been performed with a MCMC method implementing the \emph{stretched move} algorithm proposed by \citet{GW2010} and including parallel tempering as described, e.g., by \citet[][]{Benomar_2009}. The latter improved the exploration of parameter spaces containing numerous local maxima. We typically sample the posterior distributions with 20 parallel chains and 300 walkers for each chain. The initial positions of the walkers are randomly drawn from the prior distributions. The walkers are iterated over 15\,000 steps, with the 5\,000 first steps discarded as burn-in to ensure that chains are stationary.
After the sampling process, we assessed the completeness of the sampling for every parameter, by looking at the shape of the posterior distributions and by checking the correlation times. They are defined as the number of steps needed for the autocorrelation function of the chain to be divided by $e$. In our case, the correlation time is about 37, which ensures to get about 81\,000 uncorrelated sampling points.

As a result of the sampling, the median of the posterior distributions of the model parameters and the 1-$\sigma$ errors are presented in Table~\ref{mag_seismic_param}. The values of $\epsilon_{g_1}$ and $\epsilon_{g_2}$ remain quite close and the estimated $\nu_{\text{rot}}$ is coherent with the one found with the rotational modulation. We find a significant detection for the magnetic field since $\nu_B=46.8\pm3.4$\:pHz.
To verify the sensitivity of our analysis to the selected frequencies, we also performed analyses by rejecting frequencies below a 5-$\sigma$ detection level (instead of 4-$\sigma$), that is a SNR of 5.1, and frequencies which are combinations within 3-$\sigma$ error bars (instead of 2-$\sigma$), that is removing also $f_{66}, \ f_{28}, \ f_{44}, \ f_{45}, \ f_{68}$ and $f_{71}$. The determination of $\nu_B$ only slightly changes, decreased by less than 10\%, is still highly significant, and does not change the conclusion of this paper.

\section{Model spectrum with a dip}\label{dip_description}
We adapted the model proposed by \citet{Tokuno2022} and \citet{Galoy2024} to couple g-i modes with a pure inertial mode. We computed the spectrum of mixed inertial/g-i modes by finding the roots of\\
\begin{equation}
    \cot \theta_{\rm g} = \frac{q}{s_\star-s},
\end{equation}
where \\
\begin{equation}
   \theta_{\rm g} =\pi\left(\frac{s-s_{\rm g}}{\Delta s_{\rm g}}-\frac{1}{2}\right),
\end{equation}\\
\begin{equation}
    s_{\rm g} = \frac{2\nu_{\text{rot}} \Pi_0}{\sqrt{\Lambda_{\ell,m}}} \left( n+\epsilon_{g_{\ell}} \right).
\end{equation}\\
The factor $q$ is the strength of the coupling between the modes \citep[denoted $\sigma$ in][]{Galoy2024}, $s_\star$ the spin parameter of the inertial mode, and $s_g$ the spin parameters of the g-i modes. The separation $\Delta s_g$ denotes the difference of spin parameters of modes with consecutive $n$.
This model relies on six free parameters: $\Pi_0, \ \nu_{\rm rot}, \ \epsilon_{\rm g_1}, \ \epsilon_{\rm g_2}, \ s_\star$ and $q$. 

\begin{figure}[!htbp]
    \centering
    \includegraphics[width=\linewidth]{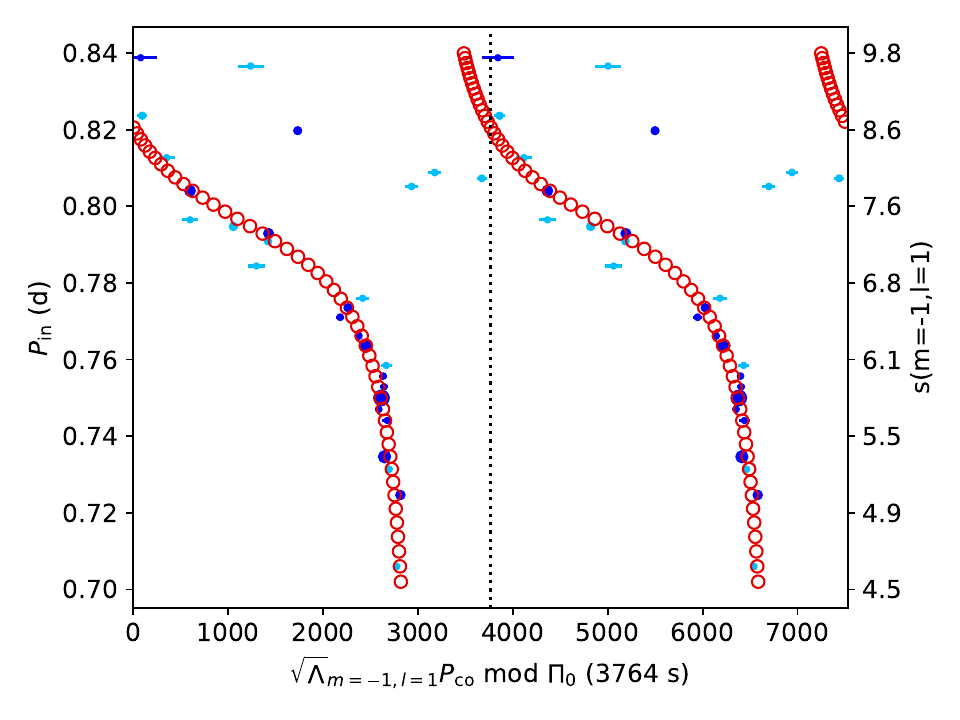}
    \includegraphics[width=\linewidth]{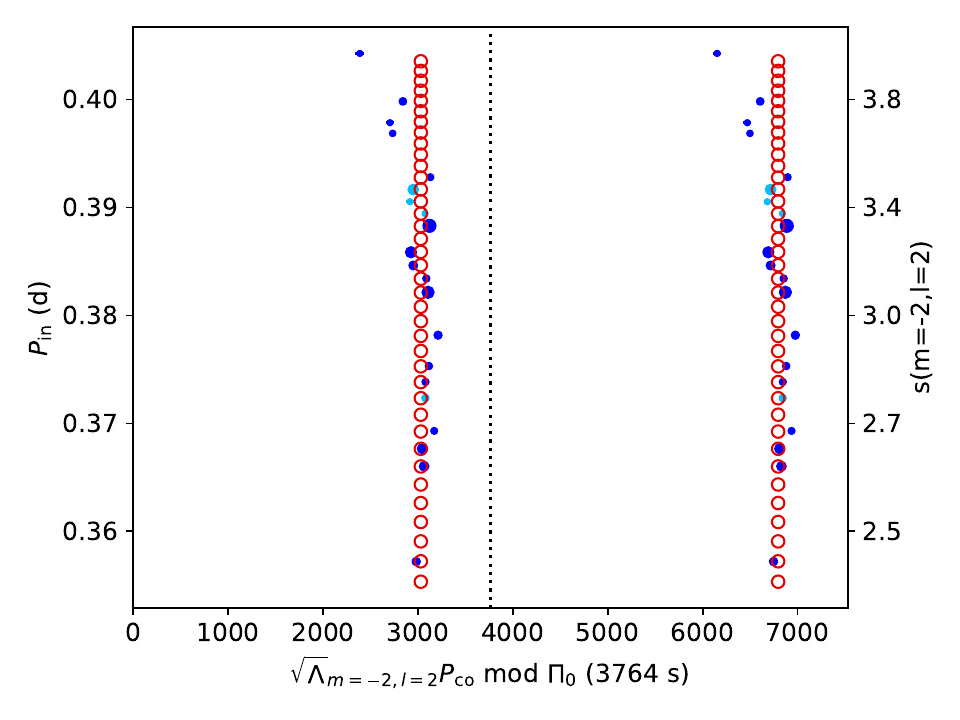}
    \caption{%Same as Fig.~\ref{fig:echelle_l12_mag} but considering a model including a dip.
    Stretched period échelle diagram of $\ell=1$ (top) and $\ell=2$ (bottom) mode frequencies. Navy blue dots: observed frequencies used for the analysis. Light blue dots: observed frequencies that are possible frequency combinations and not used for the analysis. Red circles: best-model frequencies. The size of the dots is proportional to the mode amplitude.
    }
    \label{echelle_l12_dip}
\end{figure}

We fitted this model to the observations with the method described in Sect.~\ref{ABIM}. We used the same prior distributions for the buoyancy radius, the rotation frequency, the gravity offsets and the model error parameters as those used for the magnetic model. The prior of $s_\star$ follows a uniform law over $\left[7, \ 9.5 \right]$ to cover the possible spin parameter range within which the coupling could occur, according to the oscillation spectrum of \megamite. Following \citet{Galoy2024}, we set a uniform prior over $\left[0.1, \ 2\right]$ for $q$. The prior distributions of the parameters are summarised in Table~\ref{dip_seismic_prior}.
Table \ref{dip_seismic_param} presents the median and the 1-$\sigma$ error bars of the parameters deduced from posterior distributions. Figure~\ref{echelle_l12_dip} displays the échelle diagrams of both $\ell=1$ and $\ell=2$ Kelvin modes observed and modelled frequencies. This model struggles to reproduce all the observed frequencies, contrary to the magnetic model.

\begin{table}[!htbp]
\caption{Prior distributions for the dip model}
\label{dip_seismic_prior}
\centering
\begin{tabular}{c c c}
\hline  \hline
    Parameter & Distribution & Interval \\
\hline
    $\Pi_0$ (s) & Uniform & $\left[3000, \ 5000\right]$\\
    $\nu_R \ \left( \mu \text{Hz} \right)$ & Uniform & $\left[11, \ 12 \right]$\\
    $\epsilon_{g_\ell}$ & Uniform periodic & $\left[0, \ 1\right]$ \\
    $s_\star$ & Uniform & $\left[7, \ 9.5\right]$\\
    $q$ & Uniform & $\left[0.1, \ 2\right]$\\
    $\sigma_{\ell_1}$ (s) & Uniform & $\left[0, \ 2000\right]$\\
    $\sigma_{\ell_2}$ (s) & Uniform & $\left[0, \ 1000\right]$ \\
\hline
\end{tabular}
\end{table}

\begin{table*}[!htbp]
\caption{Estimated seismic parameters for the dip model}
\label{dip_seismic_param}
\centering
\begin{tabular}{c c c c c c c c}
\hline  \hline
    $\Pi_0$ (s) & $\nu_R \ \left( \mu \text{Hz} \right)$ & $\epsilon_{g_1}$ & $\epsilon_{g_2}$ & $s_\star$ & $q$ & $\sigma_{\ell_1}$ (s) & $\sigma_{\ell_2}$ (s)\\ [0.5ex]
\hline
    $3912^{+66}_{-127} $ & $11.552^{+0.029}_{-0.067} $ & $0.082 \pm 0.324$ & $0.92 \pm 0.33$ & $7.74^{+0.38}_{-0.44}$ & $1.13^{+0.59}_{-0.38}$ & $2.78^{+3.25}_{-2.57}$ & $4.60^{+1.50}_{-9.55}$\\
\hline
\end{tabular}
\end{table*}

\section{Structure Model of \megamite}\label{app:cesam}
We computed for our study structure models calibrated on \megamite.  We calibrated the model with its buoyancy radius ($\Pi_0=3880\pm100$\:s). We used measurements reported in the \emph{Gaia} DR3 \citep{Gaia2023} for its effective temperature ($T_\mathrm{eff}=6938\pm38$\:K), it surface metallicity ($(Z/X)_s=0.0108\pm0.0035$) and its bolometric luminosity. For the latter, we find two significantly different values in the \emph{Gaia} DR3: the GSP-Phot luminosity $L/L_\sun=10.49\pm 0.90$ and the FLAME luminosity $L/L_\sun=8.68\pm0.25$. We thus made models calibrated with this two values.

Models have been computed using the \texttt{CESAM2k20} code \citep{Morel1997, Morel2008, Manchon2025} and the Optimal Stellar Models (OSM) package. Mass and age are free parameters (mainly constrained by $T_\mathrm{eff}$ and $L$).
We adopted a solar metal mixture \citep{Asplund2009} with meteoritic abundances for refractory elements from \citet{Serenelli2010}. The initial metallicity $(Z/X)_i$ is a free parameter (mainly constrained by $(Z/X)_s$), and we considered three values for the initial helium abundance $Y_i=0.25$, 0.26 and 0.27. We used opacity tables from OPAL, the OPAL2005 equation of state \citep{Rogers2002} and the nuclear reaction rates from the NACRE collaboration \citep{Angulo1999_NACRE} except for the $\rm {}^{14}N(p,\gamma){}^{15}O$ reaction, for which we used the LUNA reaction rate given in \citet{Imbriani2004}. Convection was treated using the mixing-length theory \citep{BohmVitense1958} with a mixing-length parameter $\alpha_\mathrm{MLT} = 1.77$, close to a solar calibration, $\alpha_\mathrm{MLT} = 1.0$, or $\alpha_\mathrm{MLT} = 0.5$. Models include the effect of atomic diffusion following the \citet{MichaudProffitt1993} formalism with radiative acceleration of \citet{Alecian2020}. We adopted the Montreal prescription \citep{Richer2000} for turbulent diffusion, calibrated with the parametrisation of \citet{VermaSilvaAguirre2019}. Models \rev{include} also overshooting of the convective core with three prescriptions: (i) diffusive overshoot, (ii) instant mixing overshoot, (iii) instant mixing overshoot with adiabatic extension of the core. The overshoot parameter is free, mainly calibrated to reproduce $\Pi_0$. The overshoot needed to reproduce $\Pi_0$ is typically $0.23\sim0.29H_p$ for instant mixing or $0.024\sim0.028$ for diffusive mixing. 
These values are consistent with the study of \citet{Mombarg2021}.
We also computed a fourth series of models with a fixed  overshoot of $0.1H_p$, but with a free extra uniform turbulent diffusion in the radiative zone, to build models similar to the $\gamma$ Dor models of \citet{Ouazzani2020}. We need a diffusion coefficient $D_t=600\sim750\:\mathrm{cm^2\, s^{-1}}$ to fit the observations. Here we used different overshoot or turbulent mixing prescriptions to mimic any mixing processes that would impact these layers (for example, rotational mixing).

We ended up with 72 models with masses from 1.36 to 1.56$M_\sun$ with central hydrogen ranging between 0.25 and 0.38, that is between one half and two third of the central hydrogen reservoir has been burned. We used these different models to explore the influence of modelling uncertainties on the magnetic field strength determination. The value of $Y_i$ has almost no impact, $\alpha_\mathrm{MLT}$ has a strong impact on the upper layers, the amplitude of $K_r(r)$ below the convective zone and the value of $\cal I$, but finally no impact on the measured value of magnetic field above the core. The overshoot and mixing prescription changes the shape of the Brunt-Väisälä frequency above the convective core, thus the detailed shape of $K_r(r)$, but it only weakly affects the estimated magnetic field strength in this region.

\begin{figure}[!htbp]
    \centering
    \includegraphics[width=\linewidth]{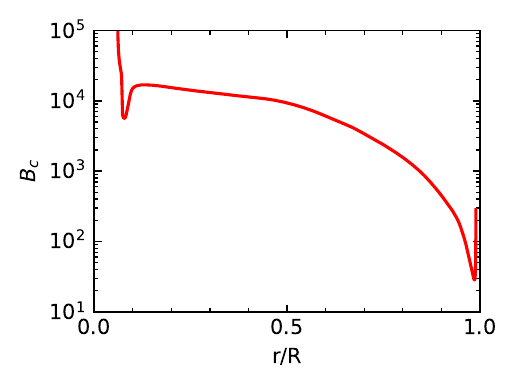}
    \caption{Critical magnetic field as a function of the radius, plotted for a representative model of \megamite\ with $Y_i=0.26$, a diffusive overshoot and calibrated of the GSP-Phot luminosity (see text).}
    \label{Bcrit}
\end{figure}

\rev{Figure \ref{Bcrit} displays the critical magnetic field profile $B_c$ in a model of \megamite \ computed for the lowest $\ell =1$ mode frequency ($f_{66}$). We verify that, near the core, the magnetic field measured in \megamite\ (about 4 kG) is lower than $B_c$. As $B_c$ increases with the frequency, the measured field remains lower than $B_c$ for all the observed modes.
We performed this computation with our 72 models and verified that a field $B_r \approx 4$\: kG remains below $B_c$ for all of them.} 

\begin{figure}[!htbp]
    \centering
    \includegraphics[width=\linewidth]{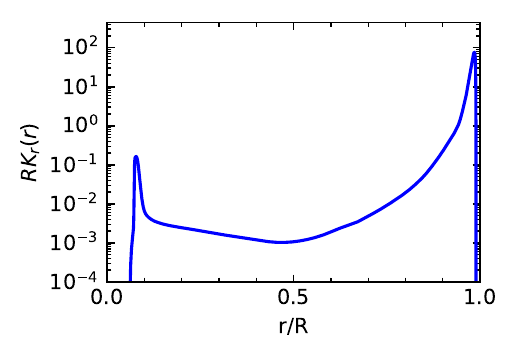}
    \caption{Magnetic weight function $K_r(r)$ adimensionalised by the stellar radius $R$ as a function of the normalised radius $r$, computed for a model of \megamite.}
    \label{K_r}
\end{figure}

\section{Extracted frequencies of \megamite}\label{app:felix}

\begin{figure*}[!ht]
    \centering
    \includegraphics[width=\textwidth]{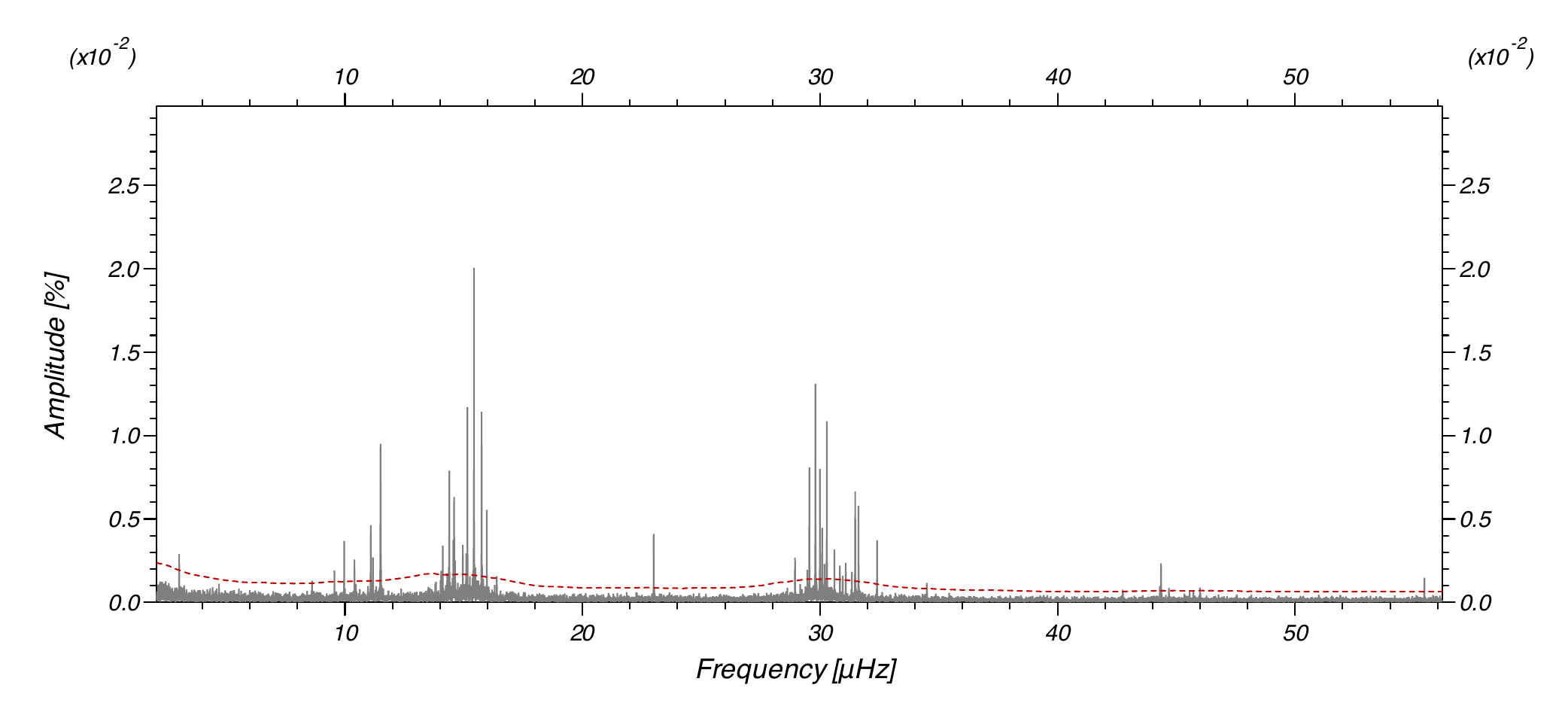}
    \caption{Oscillation spectrum of \megamite. The frequency axis is represented in $\mu$Hz and the amplitude axis is represented in~$\%$. The red dashed curve represents the 4-$\sigma$ noise threshold before pre-whitening.}
    \label{spectrum_2309579}
\end{figure*}

We performed the analysis of the light curve of \megamite\ with a pre-whitening technique using the \texttt{FELIX} code \citep{Charpinet2010, Zong2016}. We extracted frequencies with a signal-to-noise ratio (SNR) greater than 4.7, that is a 4-$\sigma$ detection level. Figure \ref{spectrum_2309579} displays the oscillation spectrum of \megamite. Table \ref{frequencies} lists the extracted frequencies along with their errors, amplitudes, and signal-to-noise ratios. Since linear combinations and/or harmonics could arise, we precise the possible linear combinations within 2-$\sigma$ error bars.
We clearly identified $\ell=1$ and $\ell=2$ Kelvin modes as groups of peaks around 15 and 30\:$\mu$Hz. We also identify a peak around 11.50\:$\mu$Hz as the surface rotation $\nurot$, as well as its harmonics at 2, 3, 4$\nurot$. During the pre-whitening process, the rotation peak has been fitted with two very close large peaks, $f_{01}$ and $f_{02}$, with different phases, presumably to take into account a modulation of its amplitude\rev{, which could be due to the finite lifetime of surface spots, latitudinal differential rotation, or even time variations of surface chemical inhomogeneities.} Near the \rev{rotation frequency}, there are a few frequencies that could be identified as Rossby modes and would require a dedicated analysis.

For the frequencies that have been modelled during the analysis, Table \ref{frequencies} shows their associated degree $\ell$ and azimuthal order $m$. We also carried out an \textit{a posteriori} identification of their radial orders $n$ considering the optimal parameters of our model. Frequencies that could be linear combinations within a 2-$\sigma$ error bar but which are well represented by our best-model are noted with a question mark next to their radial orders.
\onecolumn
\begin{longtable}{c c c c c c c c c}
\caption{Extracted frequencies of \megamite}
\label{frequencies}\\
%\centering
%\begin{longtable}{c c c c c c c c c}
\hline \hline
\ & $f \ \left(\mu \text{Hz}\right)$ & $\sigma_f \left( \text{nHz}\right)$ & Amplitude $\left( \rm ppm \right)$ & S/N & $\ell$ & $m$ & $n$ & Comments\\ 
\hline
\endfirsthead
\caption{continued.}\\
\hline
\ & $f \ \left(\mu \text{Hz}\right)$ & $\sigma_f \left( \text{nHz}\right)$ & Amplitude $\left( \rm ppm \right)$ & S/N & $\ell$ & $m$ & $n$ & Comments\\ 
\hline 
\endhead
\hline
    %$f_{29}$ & 0.56526382 & 0.6958 & $24.7$ & 6.28 & \ & \ & \ & \\
    %$f_{39}$ & 0.58552603 & 0.8939 & $19.0$ & 4.89 & \ & \ & \ & \\
    %$f_{22}$ & 0.67596878 & 0.5025 & $33.4$ & 8.70 & \ & \ & \ & \\
    %$f_{13}$ & 0.68695464 & 0.3043 & $55.1$ & 14.37 & \ & \ & \ & \\
    $f_{24}$ & 3.02154171 & 0.4020 & $29.2$ & 10.87 & \ & \ & \ & \\
    $f_{59}$ & 4.69837667 & 0.8870 & $10.2$ & 4.93 & \ & \ & \ & $f_{66}-f_{04}:0.6\sigma$\\
    $f_{46}$ & 8.61462975 & 0.5335 & $13.8$ & 8.19 & \ & \ & \ & $f_{13}-f_{49}:1.2\sigma$\\
    $f_{35}$ & 9.55865285 & 0.3598 & $21.1$ & 12.15 & \ & \ & \ & \\
    $f_{20}$ & 9.96821817 & 0.2120 & $35.4$ & 20.62 & \ & \ & \ & \\
    $f_{27}$ & 10.4014541 & 0.2974 & $25.1$ & 14.70 & \ & \ & \ & \\
    $f_{58}$ & 10.9760210 & 0.7264 & $10.4$ & 6.02 & \ & \ & \ & \\
    $f_{37}$ & 11.0792432 & 0.4050 & $18.8$ & 10.79 & \ & \ & \ & $f_{74}-f_{29}:0.3\sigma$\\
    $f_{15}$ & 11.0894796 & 0.1504 & $50.4$ & 29.06 & \ & \ & \ & $f_{41}-f_{29}:1.7\sigma$\\
    $f_{25}$ & 11.1814386 & 0.2739 & $27.6$ & 15.96 & \ & \ & \ & \\
    $f_{63}$ & 11.4760733 & 0.8350 & $9.2$ & 5.24 & \ & \ & \ & \\
    $f_{02}$ & 11.4985523 & 0.0079 & $973.4$ & 552.88 & \ & \ & \ & \\
    $f_{01}$ & 11.4988648 & 0.0076 & $1016.5$ & 577.38 & \ & \ & \ & \nurot\\
    $f_{66}$ & $13.7988610$ & $0.928$ & $8.6$ & 4.71 & $1$ & $-1$ & $117$ & \\
    $f_{57}$ & $13.8345810$ & $0.7460$ & $10.7$ & 5.86 & $1$ & $-1$ & $115?$ & $f_{04}-f_{14}:0.7\sigma$ \\
    $f_{33}$ & $14.0521910$ & $0.3550$ & $21.9$ & 12.32 & \ & \ & \ & $f_{04}-f_{06}:0.3\sigma$ \\
    $f_{22}$ & $14.1193500$ & $0.2380$ & $32.8$ & 18.40 & $1$ & $-1$ & $102$ & \\
    $f_{53}$ & $14.2420280$ & $0.6490$ & $11.9$ & 6.74 & \ & \ & \ & $f_{08}-f_{06}:1.1\sigma$\\
    $f_{43}$ & $14.3097500$ & $0.5200$ & $14.8$ & 8.41 & $1$ & $-1$ & $95?$ & $f_{39}-f_{05}:1.0\sigma$\\
    $f_{38}$ & 14.3367620 & 0.4170 & $18.5$ & 10.48 & $1$ & $-1$ & $94?$ & $f_{16}-f_{06}:0.1\sigma$\\
    $f_{47}$ & 14.3748500 & 0.5620 & $13.8$ & 7.78 & \ & \ & \ & $f_{04}-f_{03}:0.2\sigma$\\
    $f_{09}$ & $14.3949860$ & $0.1000$ & $77.6$ & 43.93 & $1$ & $-1$ & $92$ & \\
    $f_{60}$ & $14.5318290$ & $0.806$ & $9.6$ & 5.42 & \ & \ & \ & $f_{07}-f_{06}:0.8\sigma$ \\
    $f_{19}$ & $14.5636310$ & $0.2180$ & $35.6$ & 20.08 & $1$ & $-1$ & $87?$ & $f_{08}-f_{03}:1.5\sigma$\\
    $f_{12}$ & 14.5969310 & 0.1260 & $61.5$ & 34.63 & $1$ & $-1$ & $86$ & \\
    $f_{28}$ & 14.6347660 & 0.3270 & $23.7$ & 13.38 & $1$ & $-1$ & $85$ & \\
    $f_{72}$ & 14.7551170 & 0.9710 & $8.0$ & 4.50 & \ & \ & \ & $f_{56}-f_{10}:0.0\sigma$\\
    $f_{62}$ & 14.9149600 & 0.8060 & $9.6$ & 5.43 & $1$ & $-1$ & $78?$ & $f_{07}-f_{54}:0.5\sigma$\\
    $f_{23}$ & 14.9627470 & 0.2370 & $32.6$ & 18.43 & $1$ & $-1$ & $77$ & \\
    $f_{44}$ & 15.0109300 & 0.5300 & $14.6$ & 8.24 & $1$ & $-1$ & $76$ & \\
    $f_{32}$ & 15.1066030 & 0.3470 & $22.2$ & 12.61 & $1$ & $-1$ & $74$ & \\
    $f_{05}$ & 15.1565570 & 0.0650 & $118.3$ & 67.35 & $1$ & $-1$ & $73$ & \\
    $f_{64}$ & 15.2600990 & 0.8320 & $9.1$ & 5.25 & $1$ & $-1$ & $71?$ & $f_{30}-f_{52}:1.1\sigma$\\
    $f_{50}$ & 15.3161150 & 0.5820 & $12.9$ & 7.51 & $1$ & $-1$ & $70$ & \\
    $f_{54}$ & 15.3731870 & 0.6380 & $11.7$ & 6.86 & $1$ & $-1$ & $69$ & \\
    $f_{03}$ & 15.4324980 & 0.0370 & $199.5$ & 116.93 & $1$ & $-1$ & $68$ & \\
    $f_{45}$ & 15.4936900 & 0.5200 & $143$ & 8.41 & $1$ & $-1$ & $67$ & \\
    $f_{68}$ & 15.5547920 & 0.8760 & $8.5$ & 4.99 & $1$ & $-1$ & $66$ & \\
    $f_{06}$ & 15.7551690 & 0.0660 & $111.2$ & 66.02 & $1$ & $-1$ & $63$ & \\
    $f_{52}$ & 15.8252880 & 0.6090 & $12.1$ & 7.18 & $1$ & $-1$ & $62$ & \\
    $f_{14}$ & 15.9721310 & 0.1380 & $53$ & 31.70 & $1$ & $-1$ & $60$ & \\
    $f_{42}$ & 16.3945560 & 0.4770 & $14.8$ & 9.17 & $1$ & $-1$ & $55$ & \\
    $f_{17}$ & 22.9982834 & 0.1672 & $3.9$ & 26.14 & \ & \ & \ & 2\nurot\\
    $f_{49}$ & 23.0056685 & 0.5069 & $13.0$ & 8.63 & \ & \ & \ & \\
    $f_{65}$ & 28.6298020 & 0.7240 & $9.0$ & 6.04 & $2$ & $-2$ & $94$ & \\
    $f_{26}$ & 28.9481710 & 0.2420 & $26.8$ & 18.05 & $2$ & $-2$ & $89$ & \\
    $f_{71}$ & 29.0917110 & 0.7920 & $8.2$ & 5.52 & $2$ & $-2$ & $87$ & \\
    $f_{51}$ & 29.1641860 & 0.5220 & $12.5$ & 8.37 & $2$ & $-2$ & $86$ & \\
    $f_{39}$ & 29.4656660 & 0.3490 & $18.4$ & 12.53 & $2$ & $-2$ & $82$ & \\
    $f_{10}$ & 29.5517980 & 0.0830 & $77.0$ & 52.47 & $2$ & $-2$ & $81?$ & $f_{05}+f_{09}:1.8\sigma$\\
    $f_{75}$ & 29.6369050 & 0.8350 & $7.7$ & 5.24 & $2$ & $-2$ & $80?$ & $f_{32}+f_{60}:1.3\sigma$\\
    $f_{67}$ & 29.7197070 & 0.7570 & $8.5$ & 5.77 & $2$ & $-2$ & $79?$ & $f_{05}+f_{19}:0.6\sigma$\\
    $f_{04}$ & 29.8072560 & 0.0490 & $130.8$ & 88.72 & $2$ & $-2$ & $78$ & \\
    $f_{08}$ & 29.9964790 & 0.0740 & $8.5$ & 59.10 & $2$ & $-2$ & $76$ & \\
    $f_{16}$ & 30.0918700 & 0.1400 & $45.1$ & 31.15 & $2$ & $-2$ & $75$ & \\
    $f_{34}$ & 30.1868560 & 0.2950 & $21.2$ & 14.82 & $2$ & $-2$ & $74$ & \\
    $f_{07}$ & 30.2876270 & 0.0570 & $110.3$ & 76.20 & $2$ & $-2$ & $73$ & \\
    $f_{21}$ & 30.6058160 & 0.1840 & $34.2$ & 23.81 & $2$ & $-2$ & $70$ & \\
    $f_{31}$ & 30.8387980 & 0.2740 & $22.7$ & 15.96 & $2$ & $-2$ & $68$ & \\
    $f_{36}$ & 30.9602270 & 0.3200 & $19.4$ & 13.68 & $2$ & $-2$ & $67$ & \\
    $f_{30}$ & 31.0842580 & 0.2700 & $23.1$ & 16.21 & $2$ & $-2$ & $66$ & \\
    $f_{49}$ & 31.3406810 & 0.3480 & $18$ & 12.57 & $2$ & $-2$ & $64$ & \\
    $f_{11}$ & 31.4812770 & 0.0960 & $66$ & 45.75 & $2$ & $-2$ & $63$ & \\
    $f_{13}$ & 31.6211660 & 0.1120 & $56.6$ & 39.18 & $2$ & $-2$ & $62$ & \\
    $f_{18}$ & 32.4025250 & 0.1710 & $36.5$ & 25.54 & $2$ & $-2$ & $57$ & \\
    $f_{48}$ & 34.4967813 & 0.4725 & $13.1$ & 9.25 & \ & \ & \ & 3\nurot\\
    $f_{61}$ & 34.5062545 & 0.6454 & $9.6$ & 6.77 & \ & \ & \ & \\
    $f_{77}$ & 42.7489582 & 0.7814 & $6.9$ & 5.59 & \ & \ & \ & \\
    $f_{56}$ & 44.3069656 & 0.5265 & $10.7$ & 8.30 & \ & \ & \ & $f_{22}+f_{34}:1.2\sigma$\\
    $f_{29}$ & 44.3578455 & 0.2397 & $23.6$ & 18.23 & \ & \ & \ & $f_{41}-f_{15}:1.7\sigma$\\
    $f_{69}$ & 44.6887790 & 0.6609 & $8.5$ & 6.61 & \ & \ & \ & $f_{12}+f_{16}:0.0\sigma$\\
    $f_{79}$ & 45.5615112 & 0.8607 & $6.6$ & 5.08 & \ & \ & \ & $f_{06}+f_{04}:1.1\sigma$\\
    $f_{76}$ & 45.7122074 & 0.7496 & $7.6$ & 5.83 & \ & \ & \ & $f_{32}+f_{21}:0.3\sigma$\\
    $f_{70}$ & 45.9956586 & 0.7058 & $8.2$ & 6.19 & \ & \ & \ & 4\nurot\\
    $f_{74}$ & 55.4373428 & 0.7120 & $7.8$ & 6.14 & \ & \ & \ & \\
    $f_{41}$ & 55.4465469 & 0.3666 & $15.2$ & 11.93 & \ & \ & \ & $f_{15}+f_{29}:1.7\sigma$\\
    $f_{78}$ & 58.4258909 & 0.8226 & $6.7$ & 5.31 & \ & \ & \ & \\
    $f_{73}$ & 59.5550624 & 0.7023 & $7.9$ & 6.23 & \ & \ & \ & $f_{26}+f_{21}:1.4\sigma$\\
    $f_{55}$ & 59.7386007 & 0.4777 & $11.6$ & 9.15 & \ & \ & \ & $f_{10}+f_{34}:0.1\sigma$\\
\hline
\\[-0.8em]
\multicolumn{9}{l}{\parbox{0.8\linewidth}{\footnotesize \textbf{Notes.} In the first column, the frequencies are labelled according to their amplitudes. In the \emph{Comments} column, possible linear combinations within $2\sigma$ are listed in the form $f_a \pm f_b:n\sigma$ and rotation frequency and its first harmonics are noted in the form $n$ \nurot.}} \\
\end{longtable}

%\FloatBarrier
\twocolumn

\end{appendix}
\end{document}